\def\ds		{\displaystyle}
\def\d		{{\rm d}}
\def\saut	{\vskip0.5cm}

\def\up		{+}
\def\dw		{-}

\def\Pmic	{P^{\rm mic}_{m \alpha  \beta}}

\def\Pmicp	{P^{\rm mic}_{m' \alpha' \beta'}}
\def\Peff	{P^{\rm eff}_m}
\def\Peffp	{P^{\rm eff}_{m'}}

\def\Nc		{C_m}

\def\Asa	{A_m}

\def\Ess	{\varepsilon_{\rm bk}}
\def\gss	{g_{\rm bk}}
\def\EH		{\varepsilon_{\rm sh}}
\def\gH		{g_{\rm sh}}
\def\Hmic	{{\cal H}^{\rm mic}_{m  \alpha \beta}}
\def\Hsig	{{\cal H}^{\rm mac}_{m  \sigma}}
\def\Hsigp	{{\cal H}^{\rm mac}_{m' \sigma'}}

\def\Hmicp	{{\cal H}^{\rm mic}_{m'  \alpha'   \beta'}}
\def\Heff	{{\cal H}^{\rm eff}_m}
\def\Heffp	{{\cal H}^{\rm eff}_{m'}}

\def\Qmic	{{\cal P}^{\rm mac}_{m \sigma}}
\def\Qmicp	{{\cal P}^{\rm mac}_{m' \sigma'}}

\def\taumic	{\tau^{\rm mic}_{m,m'}}
\def\taueff	{\tau_{\rm eff}}

\documentclass[twocolumn,showpacs,preprintnumbers,amsmath,amssymb]{revtex4}
\usepackage{graphicx}
\usepackage{dcolumn}
\usepackage{bm}
\usepackage{color}
\begin{document}

\preprint{}

\title{
Master equation of proteins in interaction with implicit or explicit solvent.
}

\author{Olivier Collet}

\affiliation{
Institut Jean Lamour, D\'epartement 1, CNRS, Universit\'e de Lorraine \\
Boulevard des Aiguillettes BP 239, F-54506 Vandoeuvre-l\`{e}s-Nancy
}



\begin{abstract}
Theoretical studies of protein folding on lattice models relie on the 
assumption that water close to amino-acids is always in thermal 
equilibrium all along the folding pathway. Within this framework, it has 
always been considered that out-of-equilibrium properties, such as 
folding time, could be evaluated equivalently from an averaging over a 
collection of trajectories of the protein with water described either 
explicitly or through a mean-field approach.
To critically assess this hypothesis, we built a two-dimensional lattice 
model of a protein in interaction with water molecules that  can adopt a 
wide range of conformations. This microscopic description of the solvent 
has been used further to derive an effective model by averaging over all 
the degrees of freedom of the solvent. At thermal equilbrium, the two 
descriptions are rigourously equivalent, predicting the same folded 
conformation of the protein.  
The model allows exact calculations of some relaxation properties
using the master equations associated to both solvent descriptions.
The kinetic patterns associated to the folding pathways are remarkably different. In this 
work we demonstrate, that an effective description of the solvent can 
not described properly the folding pathway of a protein. The microscopic 
solvent model, that describes correctly the microscopic routes,  appears 
to be the only candidate to study folding kinetics.
\end{abstract}

 \maketitle

Proteins folding is a hot topic of the biophysics field and
the question of the mechanisms which governs its kinetics is still in debate.
The two ingredients guiding a protein towards its native structure
are the distribution of the intrachain interactions and the solvent effect\cite{Kauzmann1959}.
Thus, the solvation of hydrophobic compounds at thermal equilibrium
 has been widely studying
to extract the key role of the water in protein folding\cite{Chandler1993,Hummer1998a,Chothia1974,Lee1971,Silverstein1999,Becker2006}.

However, in most of the works using lattice models for protein, 
the effect of the solvent is usually taken into account
by a temperature independent, structure-less, parameter which simply increases the 
strength of the some intrachain contact\cite{Shakhnovich1990a,Lau1989}.
The lattice model using such couplings also provided numerous kinetics works 
using Monte Carlo simulations \cite{Gutin1995b,Gutin1995,Chan1998,Collet2003b} 
or evolutions of the master equation\cite{Chan1994,Zwanzig1995,Cieplak1998,Pitard2000,Kachalo2006,Zhou2008,Collet2008}.
In these works, 
as the potential associated to each protein structure results from an average over the degrees of freedom,
the kinetics is guided by transition rates between chain conformations
in interaction with an effective solvent whose mean energy does not depend on the temperature.
However, numerous experimental results showing the importance of the relaxation of the first shell solvent 
on the folding kinetics 
illuminate the importance 
of the degrees of freedom of the first shell
solvent for the fast kinetics of folding\cite{Zanotti2008,Frauenfelder2006,Lubchenko2005,Shenogina2008}

In a few recent works, the contribution of the solvent effect on the configurational
Hamiltonian becomes temperature dependent 
because they result from an average over the water configurations. 
In these models, the solvent around the proteins is modeled by its energy spectra which takes into account of the
formation or breakage of the water hydrogen bonds.
Such an approach gave an explanation of cold denaturation\cite{DeLosRios2000,Collet2001}.

Recently, we calculated the kinetics of the folding of a protein model in 
interaction with implicit water model\cite{Collet2011} where the role of the hydrogen bonds of the
first shell solvent was taken into account.

Here, we compare the results obtained for the kinetics of folding of protein where each chain structure is
in interaction with a highly degenerated microscopic solvent in one hand 
and with the equivalent effective solvent in an other hand.
A micro-state of the system "protein-solvent" is given by 
the conformation of the protein 
and the position of the atoms of water (the solvent configuration).
We calculated the evolution in time of the probability of occurrence of each
protein-solvent micro-state toward the native structure of the protein,
using a master equation approach  and
starting far from equilibrium.
\begin{figure}[hbtp]
\centerline{\includegraphics[width=3.4in]{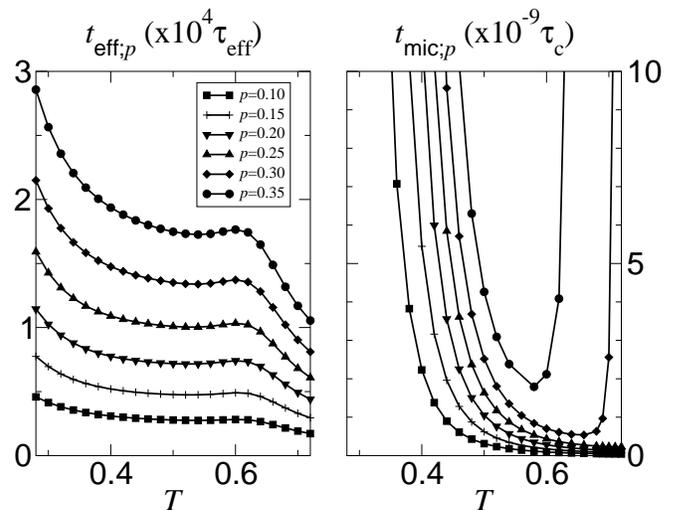}}
\vskip-0.5cm
\caption{\label{folding}
Time after which the probability of occurrence of the native state equals some values $p$
as function of the temperature starting from an equiprobability of each protein-solvent configurations.
Calculations are performed with the explicit (left) and the implicit (right) solvent models.
}
\end{figure}
The waiting times to observe the native structure with a probability $p$, noted 
$t_{{\rm mic} ; p}$ are calculated
as functions of the temperature for $p$ varying from 0.10 to 0.30 by steps of 0.05.

On an other hand, the effective solvent model is introduced by
integrating out the water degrees of freedom of the same solvent model
and by computing the free energy of hydration of
each protein conformation.
Simulations start from the equivalent initial condition
to that stated for the microscopic model calculation.
The evolution in time of the probabilities of the protein conformations 
is also computed using a master equation and 
the waiting times, $t_{{\rm eff} ; p}$, to observe the native structure with a probability $p$
are also calculated under the same conditions.

The curve of $t_{{\rm mic} ; p}(T)$ and $t_{{\rm eff} ; p}(T)$, 
shown in fig.\ref{folding},  present clearly different shapes and orders of magnitude.

\saut

{\bf Model.}\hskip0.5cm
  The protein is modeled by a self avoiding walk chain whose the twelve beads are 
  positioned on the nods of a  two-dimensional lattice.
  The number of intrachain contacts of the chain conformation $m$ is
$\Nc = \sum_{i \ge j+3} \Delta_{ij}^{(m)}$ where
$\Delta_{ij}^{(m)}=1$, if the monomers $i$ and $j$ are first neighbors 
on the lattice, and 0 otherwise.
  The accessible surface area to the solvent is $\Asa = 2N+2-2\Nc$\cite{Collet2001}.

The bulk contribution is taken as a mean effect which increases
with respect of the number of intrachain contacts and
the first shell contribution increases with the accessible surface area to the solvent.
One configuration ($\beta=0$) of the first shell is well ordered
and the other are disorganized.

The Hamiltonian of the micro-state where the protein is in conformation $m$, 
the first shell in configuration $\beta$
and the bulk in structure $\alpha$ is:
$$\Hmic =  \sum_{i > j} B_{ij} \ \Delta_{ij}^{(m)} 
+	2 \Nc \Ess
+	\sigma(\beta) \Asa \EH
$$
$$  {\rm for} \ 1 \le \alpha \le  \gss^{2 \Nc}
\quad {\rm and} \quad
\left\{
\begin{array}{l}
\sigma(0) = 0 \\
\sigma(\beta)=1 \ {\rm if} \ 1 \le \beta \le \gH^{\Asa} 
\end{array}
\right.
$$
The intrachain couplings $B_{ij}$ between monomers $i$ and $j$ are drawn at random
from a Gaussian distribution centered on $B_0 = -2$ with standard deviation equals 1 \cite{Shakhnovich1990a}. 
The values of the solvent parameters are ranked as follow 
$\Ess=0.4$, $\EH=0.8$, $\gss=3.3$ and $\gH=3.5$\cite{Collet2011}.%

The Hamiltonian of each protein structure $m$ takes two values following that of $\beta$.
The ground state, noted ($\dw$), is associated to a value of $\sigma=0$ (for $\beta=0$) 
and the exited state, ($\up$), to $\sigma=1$ (for $\beta > 0$).
The Hamiltonian and the degeneracy of the macro-states $(m\sigma)$ are~:
$$
\Hsig = \sum_{i > j} B_{ij} \ \Delta_{ij}^{(m)} + 2 \Nc \Ess + \sigma \Asa \EH \nonumber 
$$
\vskip-0.4cm
$$
g_{m\sigma}  = 
\sum_{\alpha=1}^{\gss^{2 \Nc}} \sum_{\beta=0}^{\gH^{\Asa}} \delta(\sigma - \sigma(\beta))  = 
\gH^{\sigma  \Asa} \ \gss^{2\Nc} \nonumber 
$$
where $\delta(x) = 1$ if $x=0$ and 0 otherwise.

%
Then, the partition function may be written as~:
$$\begin{array}{lll}
Z(T)  & =       \ds \sum_m \sum_{\alpha=1}^{\gss^{ 2 \Nc}} \sum_{\beta=0}^{ \gH^{\Asa}} 
        \exp\left(- \frac{\Hmic}{T} \right)  
      & \ \ {\rm (micro-states)}    \nonumber \\
      & = \ds \sum_m \sum_{\sigma=0}^1 g_{m\sigma} 
      \exp\left(- \frac{\Hsig}{T} \right)  
      & \ \ {\rm (macro-states)}    \nonumber \\
      & = \ds \sum_m \exp \left(- \frac{\Heff(T)}{T} \right) 
      & \ \ {\rm (effective)} \nonumber
\end{array}$$

The temperature dependent effective Hamiltonian of conformation $m$,
where  the solvent degrees of freedom have been integrating out, is given by~:

$$\Heff = 
- T \ln \sum_{\alpha=1}^{\gss^{ 2 \Nc}} \sum_{\beta=0}^{ \gH^{\Asa}} 
	\exp\left(- \frac{\Hmic}{T} \right) $$
that is to say,
$$\begin{array}{ll}
\ds \Heff = & \ds \sum_{i > j} B_{ij} \ \Delta_{ij}^{(m)} + 2 \Nc \Ess  \\
	    & \ds  - T \ \ln\left[\gss^{2 \Nc} \left(1+\gH^{A_m} \ 
		\exp \left(-\frac{A_m \EH} {T}\right) \right)\right]
\end{array}$$

The effective Hamiltonian has been written to satisfy to the partition function
which determines the properties of the system at equilibrium but
out-of-equilibrium, this function  have no more  a clear physical meaning.

The probability of occurrence of the conformation $m$ at equilibrium is~:
$P_m^{\rm eq} =  \exp\left( -{\Heff}/{T}\right)/Z(T)$.
The native conformation is the structure of the largest probability
determined by a full enumeration of the conformational space.

\saut

{\bf Dynamics of the microscopic solvent model.}\hskip0.5cm
The out-of-equilibrium probability of occurrence of the micro-state 
$(m \alpha \beta)$ at time $t$ is denoted $\Pmic(t)$.
It evolves following the master equation\cite{Collet2011}:
$$\frac{\d \Pmic(t)}{\d t} = \sum_{m' \alpha' \beta'} X_{m \alpha \beta , m' \alpha' \beta'} \Pmicp(t)$$
where 
$$
X_{m \alpha \beta , m' \alpha' \beta'} = \frac{V^{(0)}_{m m'}}{\taumic} 
a_T(\Hmic , \Hmicp)
$$
is the microscopic transition rate between the configurations $(m' \alpha' \beta'$ to $(m \alpha \beta$
and the diagonal terms are:
$$
X_{m \alpha \beta , m \alpha \beta} = - \sum_{m' \alpha' \beta'}
X_{m' \alpha' \beta' , m \alpha \beta} 
$$
with $V^{(0)}_{m m'}=1$ and $\taumic = \tau_c$ if the chain structures $m$ and $m'$ are connected by a one-monomer move
and $V^{(0)}_{m m}=1$ and $\tau_{mm}^{\rm mic}=\tau_s$ for the solvent modifications which 
keep the chain structure unchanged.
The acceptance function, $a_T(x,x') = [1 + \exp((x-x')/T)]^{-1}$, satisfies to the microscopic detailed balance condition.
These probabilities evolves using an Euler algorithm \cite{Cieplak1998}~:
$$\Pmic(t+\delta t) = \Pmic(t) + \delta t \sum_{m' \alpha' \beta'} 
X_{m \alpha \beta , m' \alpha' \beta'} \Pmicp(t)$$

The out-of-equilibrium probability of occurrence of the macro-state $(m \sigma)$,
at time $t$ is~:
$$\Qmic(t) = \sum_{\alpha \beta} \Pmic(t) \ \delta(\sigma-\sigma(\beta))$$
Then, at time $t + \delta t$, it becomes:
$$\begin{array}{ll}
\Qmic(t+\delta t) & = \ds \sum_{\alpha \beta} \Pmic(t + \delta t) \ \delta(\sigma-\sigma(\beta)) \\
  & \ds = \sum_{\alpha \beta} \Pmic(t) \delta(\sigma-\sigma(\beta)) \\
  &\hskip-1.3cm \ds + \sum_{\alpha, \beta} \delta t \sum_{m' \alpha' \beta'} 
    X_{m \alpha \beta , m' \alpha' \beta'} \Pmicp(t)  \ \delta(\sigma-\sigma(\beta)) 
\end{array}$$

Using the following equality
$\sum_{\sigma'} \sum_{\alpha' \beta'} \delta(\sigma'-\sigma(\beta')) f_{\beta'}
= \sum_{\alpha' \beta'} f_{\beta'}$, it comes:   
$$\begin{array}{ll}    
 \Qmic(t+\delta t)  &	\ds = \sum_{\alpha \beta} \Pmic(t) \delta(\sigma-\sigma(\beta)) \\
  &\hskip-1.3cm \ds + \delta t  \sum_{m'} \sum_{\sigma'} \sum_{\alpha' \beta'} \sum_{\alpha, \beta}
    \delta(\sigma'-\sigma(\beta')) \delta(\sigma-\sigma(\beta)) \\
  &\hskip-1.3cm  
  \frac{V^{(0)}_{m m'}}{\taumic} a_T(\Hmic , \Hmicp)  \Pmicp(t)  \  \\
& \\
  & \ds =  \Qmic(t)+  \delta t \sum_{m'} \sum_{\sigma'}  
    Y_{m \sigma, m' \sigma'} \Qmicp  
\end{array}$$
with
\begin{equation}\label{Yrate}
Y_{m \sigma, m' \sigma'} = 
g_{m \sigma}\  \frac{V^{(0)}_{m m'}}{\taumic} 
a_T(\Hsig, \Hsigp)
\end{equation}

The increment of time is chosen as 
$\delta t = 1 / \max_{m \sigma}\{Y_{m \sigma, m \sigma}\} \ll 1$
in order to maintain the sum of the probabilities equals to 1.
The form of the transition rates $Y_{m \sigma, m' \sigma'}$ implies that
the probability distribution converges to 
$\Qmic(t) \rightarrow g_{m \sigma} \exp(-\Hsig /T) / Z(T)$.

The simulations starts with the initial condition : 
$\Qmic(0) = g_{m \sigma} / \sum_{m \sigma} g_{m \sigma}$
which set the same initial weight to any micro-state.

The time needed to observe the native structure, with a probability $p$ is noted
$t_{{\rm mic};p}$ or in other words~: 
$\sum_{\sigma=0}^1 {\cal P}^{\rm mac}_{{\rm Nat},\sigma} (t_{{\rm mic};p}) = p$.

\saut

{\bf Dynamics of the effective solvent model.}\hskip0.5cm
Consider now the evolution of the probability of occurrence, $\Peff(t)$ of the chain structure $m$
in interaction with an effective solvent starting with the initial effective
probability $\Peff(0) = \sum_{\sigma} \Qmic(0)$.

 at time $t=0$.
The effective probabilities evolve following the master equation~:
$$\frac{\d \Peff(t)}{\d t} = \sum_{m'} V_{mm'} \Peffp(t)$$
where 
$V_{mm'} = V(m' \rightarrow m)$ is the transition rate from conformations 
$m'$ to $m$.
In order to satisfy to the condition of the convergence towards the equilibrium
probability distribution, a solution for the rate is~:
\begin{equation}\label{Vrate}
V_{mm'} = \frac{V^{(0)}_{m m'}}{\taueff} a_T(\Heff , \Heffp)
\end{equation}
where $\taueff$ is the effective time associated to a chain move.
Defining $V_{mm}= - \sum_{m'\ne m} V_{m'm}$ and 
using the Euler algorithm
the evolution equation reads:
$$
\Peff(t + \delta t_{\rm eff} ) =\Peff(t) +\delta t_{\rm eff} \sum_{m'} V_{mm'} \Peffp (t)
$$
with $\delta t_{\rm eff} = 1/\max_m\{V_{mm}\}$.
Obviously, the probability distribution tends towards $\Peff(t) \rightarrow P_m^{\rm eq}$.
The probability of the native structure reaches $p$ at a time, denoted by $t_{{\rm eff};p}$.

In contrast with the previous approach, the solvent degrees of freedom are
integrated first, here, and afterwards the transition rates are calculated using 
the effective potential.
This is the procedure usually applied in lattice model of protein
where the attractive term between monomers results indeed of an average
of the solvent degrees of freedom.

\saut

{\bf Discussion.}\hskip0.5cm
We address now, the question of a possible equivalence between both descriptions 
after rewriting the effective characteristic time of the transition between two chain structures
would be a, time independent function of the 
parameters associated to the two connected  chain conformations.
At this purpose, we require to satisfy the following equality for every protein conformation:
$$
  \ds \Peff(t) = {\cal P}_{m \dw}^{\rm mac}(t) + {\cal P}_{m \up}^{\rm mac}(t) $$ 
$$
\Rightarrow 
\frac{\d P_m^{\rm eff}(t)} {\d t} =
\frac{\d {\cal P}_{m \dw}^{\rm mac}(t)} {\d t} +
\frac{\d {\cal P}_{m \up}^{\rm mac}(t)} {\d t} $$
$$
\begin{array}{ll}
\Rightarrow &
\ds \sum_{m'} \frac{V_{mm'}^{(0)}}{\taueff} a_T(\Heff, \Heffp) \Peffp (t) =
\\
 & \ds \sum_{m'} \sum_{\sigma'} \sum_{\sigma} g_{m \sigma} \frac{V_{mm'}^{(0)}}{\taumic}
  a_T(\Hsig, \Hsigp) \Qmicp(t)
\end{array}$$

As only chain moves may be considered, the above equation leads to~:
$$
\ds \taueff = \tau_c \frac
{a_T(\Heff, \Heffp) \Peffp (t)}
{\sum_{\sigma'} \sum_{\sigma}  g_{m \sigma}   a_T(\Hsig, \Hsigp) \Qmicp(t)}$$
where it appears clearly that $\taueff$ depends on the time in this equation.

\vskip0.3cm

{\it
This definitively proves that the kinetics of folding of the proteins
can not be understood using protein-solvent models where the
degrees of freedom of water have been integrating out in a
conformational free energy of solvation.
}

\vskip0.3cm

However, we mention that some relations may be found for extreme temperatures.
At very low temperature, as
$\Heff \rightarrow  {\cal H}_{m -}^{\rm mac}- T \ln g_{m -}$, and as
we may assume that the exited states always have a nil non-equilibrium probability,
the above equation leads to
an effective characteristic time, only depending on the ground states of the two connected 
chain conformations:
$$
\tau_{\rm eff}^{mm'} = 
\frac{1+\exp(\Delta {\cal H}_{mm'}^-/T)}
{g_{m-} + g_{m' -} \exp(\Delta {\cal H}_{mm'}^-/T)} 
\tau_c 
$$
with $\Delta {\cal H}_{mm'}^- = {\cal H}_{m -}^{\rm mac} - {\cal H}_{m' -}^{\rm mac}$.
Putting this results into eq.\ref{Vrate}, it comes 
$V_{mm'}  = Y_{m \dw, m' \dw}$ and assuming that only the ground states of the protein
structures are visited, the two kinetics becomes equivalent.
In a similar way, at very high temperature (under which the protein is unfolded), as
$\Heff \rightarrow  {\cal H}_{m +}^{\rm mac} - T \ln g_{m +}$
and as we may assume that only the excited states have a non nil probability to
occur, an equivalent relation between both characteristic time (where the - are replaced by some +)
may be found which leads
to $V_{mm'}  = Y_{m \up, m' \up}$.
In both extreme temperature cases, it is possible to rewrite effective kinetics equations
equivalent to the microscopic ones but that is not feasible at medium temperature.

\saut
{\bf Conclusion.}\hskip0.5cm
Waiting times to observe some proportions of folded proteins
have been calculated using a microscopic description of the
solvent and the equivalent mean effect on the chain conformation weights.
In both cases, the evolution of the system depends on the ratio of the difference 
of (free) energies, induced by the attempted moves, over the temperature.

In the first simulations, the (huge) configurational space 
is composed of all the protein and solvent micro-states.
The result of the acceptance function of move depends on the energy associated
to the microscopic configurations.
In other words, the protein and the solvent evolves by performing structural changes
between microscopic realizations of the system but
the calculations converge slowly towards the equilibrium distribution of
the protein conformation, as the value of $\delta t$ is very small.

In the latter, the conformational space is smaller but the folding 
takes places in an "free energy" landscape.
However, for not too small temperature, the solvation entropy contribution to 
the values of the effective Hamiltonian leads to
free energy values smaller than the ground state energy for each protein structures.
As a consequence, the folding takes place in a conformational space in which the values
of the effective Hamiltonian are not associated to a physical realization.
Here, the simulations converge very fast towards equilibrium 
(as $\delta t_{\rm eff} \ll \delta t$) but... by following non-physical routes.

As a consequence, an microscopic solvent model is the only good candidate to 
study the out-of-equilibrium folding of proteins and the effective solvent
may only be restricted to the study of equilibrium properties.

%
%
%

\begin{thebibliography}{0}
\expandafter\ifx\csname natexlab\endcsname\relax\def\natexlab#1{#1}\fi
\expandafter\ifx\csname bibnamefont\endcsname\relax
  \def\bibnamefont#1{#1}\fi
\expandafter\ifx\csname bibfnamefont\endcsname\relax
  \def\bibfnamefont#1{#1}\fi
\expandafter\ifx\csname citenamefont\endcsname\relax
  \def\citenamefont#1{#1}\fi
\expandafter\ifx\csname url\endcsname\relax
  \def\url#1{\texttt{#1}}\fi
\expandafter\ifx\csname urlprefix\endcsname\relax\def\urlprefix{URL }\fi
\providecommand{\bibinfo}[2]{#2}
\providecommand{\eprint}[2][]{\url{#2}}

\end{thebibliography}


\begin{thebibliography}{27}
\expandafter\ifx\csname natexlab\endcsname\relax\def\natexlab#1{#1}\fi
\expandafter\ifx\csname bibnamefont\endcsname\relax
  \def\bibnamefont#1{#1}\fi
\expandafter\ifx\csname bibfnamefont\endcsname\relax
  \def\bibfnamefont#1{#1}\fi
\expandafter\ifx\csname citenamefont\endcsname\relax
  \def\citenamefont#1{#1}\fi
\expandafter\ifx\csname url\endcsname\relax
  \def\url#1{\texttt{#1}}\fi
\expandafter\ifx\csname urlprefix\endcsname\relax\def\urlprefix{URL }\fi
\providecommand{\bibinfo}[2]{#2}
\providecommand{\eprint}[2][]{\url{#2}}

\bibitem[{\citenamefont{Kauzmann}(1959)}]{Kauzmann1959}
\bibinfo{author}{\bibfnamefont{W.}~\bibnamefont{Kauzmann}},
  \bibinfo{journal}{Adv. Protein Chem.} \textbf{\bibinfo{volume}{14}},
  \bibinfo{pages}{1} (\bibinfo{year}{1959}).

\bibitem[{\citenamefont{Chandler}(1993)}]{Chandler1993}
\bibinfo{author}{\bibfnamefont{D.}~\bibnamefont{Chandler}},
  \bibinfo{journal}{Phys. Rev. E} \textbf{\bibinfo{volume}{48}},
  \bibinfo{pages}{2898} (\bibinfo{year}{1993}).

\bibitem[{\citenamefont{Hummer et~al.}(1998)\citenamefont{Hummer, Garde,
  Garcia, Paulaitis, and Pratt}}]{Hummer1998a}
\bibinfo{author}{\bibfnamefont{G.}~\bibnamefont{Hummer}},
  \bibinfo{author}{\bibfnamefont{S.}~\bibnamefont{Garde}},
  \bibinfo{author}{\bibfnamefont{A.~E.} \bibnamefont{Garcia}},
  \bibinfo{author}{\bibfnamefont{M.~E.} \bibnamefont{Paulaitis}},
  \bibnamefont{and} \bibinfo{author}{\bibfnamefont{L.~R.} \bibnamefont{Pratt}},
  \bibinfo{journal}{J. Phys. Chem. B} \textbf{\bibinfo{volume}{102}},
  \bibinfo{pages}{10469} (\bibinfo{year}{1998}).

\bibitem[{\citenamefont{Chothia}(1974)}]{Chothia1974}
\bibinfo{author}{\bibfnamefont{C.}~\bibnamefont{Chothia}},
  \bibinfo{journal}{Nature} \textbf{\bibinfo{volume}{248}},
  \bibinfo{pages}{338} (\bibinfo{year}{1974}).

\bibitem[{\citenamefont{Lee and Richards}(1971)}]{Lee1971}
\bibinfo{author}{\bibfnamefont{B.}~\bibnamefont{Lee}} \bibnamefont{and}
  \bibinfo{author}{\bibfnamefont{F.~M.} \bibnamefont{Richards}},
  \bibinfo{journal}{J. Mol. Biol.} \textbf{\bibinfo{volume}{55}},
  \bibinfo{pages}{379} (\bibinfo{year}{1971}).

\bibitem[{\citenamefont{Silverstein et~al.}(1999)\citenamefont{Silverstein,
  Haymet, and Dill}}]{Silverstein1999}
\bibinfo{author}{\bibfnamefont{K.~A.~T.} \bibnamefont{Silverstein}},
  \bibinfo{author}{\bibfnamefont{A.~D.~J.} \bibnamefont{Haymet}},
  \bibnamefont{and} \bibinfo{author}{\bibfnamefont{K.~A.} \bibnamefont{Dill}},
  \bibinfo{journal}{J. Chem. Phys.} \textbf{\bibinfo{volume}{111}},
  \bibinfo{pages}{8000} (\bibinfo{year}{1999}).

\bibitem[{\citenamefont{Becker and Collet}(2006)}]{Becker2006}
\bibinfo{author}{\bibfnamefont{J.-P.} \bibnamefont{Becker}} \bibnamefont{and}
  \bibinfo{author}{\bibfnamefont{O.}~\bibnamefont{Collet}},
  \bibinfo{journal}{Journal of Molecular Structure. Theochem}
  \textbf{\bibinfo{volume}{774}}, \bibinfo{pages}{23} (\bibinfo{year}{2006}).

\bibitem[{\citenamefont{Shakhnovich and Gutin}(1990)}]{Shakhnovich1990a}
\bibinfo{author}{\bibfnamefont{E.~I.} \bibnamefont{Shakhnovich}}
  \bibnamefont{and} \bibinfo{author}{\bibfnamefont{A.~M.} \bibnamefont{Gutin}},
  \bibinfo{journal}{Nature} \textbf{\bibinfo{volume}{346}},
  \bibinfo{pages}{773} (\bibinfo{year}{1990}).

\bibitem[{\citenamefont{Lau and Dill}(1989)}]{Lau1989}
\bibinfo{author}{\bibfnamefont{K.~F.} \bibnamefont{Lau}} \bibnamefont{and}
  \bibinfo{author}{\bibfnamefont{K.~A.} \bibnamefont{Dill}},
  \bibinfo{journal}{Macromolecules} \textbf{\bibinfo{volume}{22}},
  \bibinfo{pages}{3986} (\bibinfo{year}{1989}).

\bibitem[{\citenamefont{Gutin et~al.}(1995{\natexlab{a}})\citenamefont{Gutin,
  Abkevich, and Shakhnovich}}]{Gutin1995b}
\bibinfo{author}{\bibfnamefont{A.~M.} \bibnamefont{Gutin}},
  \bibinfo{author}{\bibfnamefont{V.~I.} \bibnamefont{Abkevich}},
  \bibnamefont{and} \bibinfo{author}{\bibfnamefont{E.~I.}
  \bibnamefont{Shakhnovich}}, \bibinfo{journal}{Proc. Natl. Acad. Sci. USA}
  \textbf{\bibinfo{volume}{92}}, \bibinfo{pages}{1282}
  (\bibinfo{year}{1995}{\natexlab{a}}).

\bibitem[{\citenamefont{Gutin et~al.}(1995{\natexlab{b}})\citenamefont{Gutin,
  Abkevich, and Shakhnovich}}]{Gutin1995}
\bibinfo{author}{\bibfnamefont{A.~M.} \bibnamefont{Gutin}},
  \bibinfo{author}{\bibfnamefont{V.~I.} \bibnamefont{Abkevich}},
  \bibnamefont{and} \bibinfo{author}{\bibfnamefont{E.~I.}
  \bibnamefont{Shakhnovich}}, \bibinfo{journal}{Biochemistry}
  \textbf{\bibinfo{volume}{34}}, \bibinfo{pages}{3066}
  (\bibinfo{year}{1995}{\natexlab{b}}).

\bibitem[{\citenamefont{Chan and Dill}(1998)}]{Chan1998}
\bibinfo{author}{\bibfnamefont{H.~S.} \bibnamefont{Chan}} \bibnamefont{and}
  \bibinfo{author}{\bibfnamefont{K.~A.} \bibnamefont{Dill}},
  \bibinfo{journal}{Proteins Struct. Funct. Genet.}
  \textbf{\bibinfo{volume}{30}}, \bibinfo{pages}{2} (\bibinfo{year}{1998}).

\bibitem[{\citenamefont{Collet}(2003)}]{Collet2003b}
\bibinfo{author}{\bibfnamefont{O.}~\bibnamefont{Collet}},
  \bibinfo{journal}{Phys. Rev. E} \textbf{\bibinfo{volume}{67}},
  \bibinfo{pages}{061912} (\bibinfo{year}{2003}).

\bibitem[{\citenamefont{Chan and Dill}(1994)}]{Chan1994}
\bibinfo{author}{\bibfnamefont{H.~S.} \bibnamefont{Chan}} \bibnamefont{and}
  \bibinfo{author}{\bibfnamefont{K.~A.} \bibnamefont{Dill}},
  \bibinfo{journal}{J. Chem. Phys.} \textbf{\bibinfo{volume}{100}},
  \bibinfo{pages}{9238} (\bibinfo{year}{1994}).

\bibitem[{\citenamefont{Zwanzig}(1995)}]{Zwanzig1995}
\bibinfo{author}{\bibfnamefont{R.}~\bibnamefont{Zwanzig}},
  \bibinfo{journal}{Proc. Natl. Acad. Sci. USA} \textbf{\bibinfo{volume}{92}},
  \bibinfo{pages}{9801} (\bibinfo{year}{1995}).

\bibitem[{\citenamefont{Cieplak et~al.}(1998)\citenamefont{Cieplak, Henkel,
  Karbowski, and Banavar}}]{Cieplak1998}
\bibinfo{author}{\bibfnamefont{M.}~\bibnamefont{Cieplak}},
  \bibinfo{author}{\bibfnamefont{M.}~\bibnamefont{Henkel}},
  \bibinfo{author}{\bibfnamefont{J.}~\bibnamefont{Karbowski}},
  \bibnamefont{and} \bibinfo{author}{\bibfnamefont{J.}~\bibnamefont{Banavar}},
  \bibinfo{journal}{Phys. Rev. Lett.} \textbf{\bibinfo{volume}{80}},
  \bibinfo{pages}{3654} (\bibinfo{year}{1998}).

\bibitem[{\citenamefont{Pitard and Orland}(2000)}]{Pitard2000}
\bibinfo{author}{\bibfnamefont{E.}~\bibnamefont{Pitard}} \bibnamefont{and}
  \bibinfo{author}{\bibfnamefont{H.}~\bibnamefont{Orland}},
  \bibinfo{journal}{Europhys. Lett.} \textbf{\bibinfo{volume}{49}},
  \bibinfo{pages}{169} (\bibinfo{year}{2000}).

\bibitem[{\citenamefont{Kachalo et~al.}(2006)\citenamefont{Kachalo, Hsiao-Mei,
  and Liang}}]{Kachalo2006}
\bibinfo{author}{\bibfnamefont{S.}~\bibnamefont{Kachalo}},
  \bibinfo{author}{\bibfnamefont{L.}~\bibnamefont{Hsiao-Mei}},
  \bibnamefont{and} \bibinfo{author}{\bibfnamefont{J.}~\bibnamefont{Liang}},
  \bibinfo{journal}{Phys. Rev. Lett.} \textbf{\bibinfo{volume}{96}},
  \bibinfo{pages}{058106} (\bibinfo{year}{2006}).

\bibitem[{\citenamefont{Zhou}(2008)}]{Zhou2008}
\bibinfo{author}{\bibfnamefont{H.}~\bibnamefont{Zhou}}, \bibinfo{journal}{J.
  Chem. Phys.} \textbf{\bibinfo{volume}{128}}, \bibinfo{pages}{195104}
  (\bibinfo{year}{2008}).

\bibitem[{\citenamefont{Collet}(2008)}]{Collet2008}
\bibinfo{author}{\bibfnamefont{O.}~\bibnamefont{Collet}}, \bibinfo{journal}{J.
  Chem. Phys.} \textbf{\bibinfo{volume}{129}}, \bibinfo{pages}{155101}
  (\bibinfo{year}{2008}).

\bibitem[{\citenamefont{Zanotti et~al.}(2008)\citenamefont{Zanotti, Gibrat, and
  Bellisent-Funel}}]{Zanotti2008}
\bibinfo{author}{\bibfnamefont{J.~M.} \bibnamefont{Zanotti}},
  \bibinfo{author}{\bibfnamefont{G.}~\bibnamefont{Gibrat}}, \bibnamefont{and}
  \bibinfo{author}{\bibfnamefont{M.~C.} \bibnamefont{Bellisent-Funel}},
  \bibinfo{journal}{Phys. Chem. Chem. Phys.} \textbf{\bibinfo{volume}{10}},
  \bibinfo{pages}{4865} (\bibinfo{year}{2008}).

\bibitem[{\citenamefont{Frauenfelder et~al.}(2006)\citenamefont{Frauenfelder,
  Fenimore, Chen, and McMahon}}]{Frauenfelder2006}
\bibinfo{author}{\bibfnamefont{H.}~\bibnamefont{Frauenfelder}},
  \bibinfo{author}{\bibfnamefont{P.}~\bibnamefont{Fenimore}},
  \bibinfo{author}{\bibfnamefont{G.}~\bibnamefont{Chen}}, \bibnamefont{and}
  \bibinfo{author}{\bibfnamefont{B.}~\bibnamefont{McMahon}},
  \bibinfo{journal}{Proc., Natl., Acad., Sci., USA}
  \textbf{\bibinfo{volume}{103}}, \bibinfo{pages}{15469}
  (\bibinfo{year}{2006}).

\bibitem[{\citenamefont{Lubchenko et~al.}(2005)\citenamefont{Lubchenko,
  Wolynes, and Frauenfelder}}]{Lubchenko2005}
\bibinfo{author}{\bibfnamefont{V.}~\bibnamefont{Lubchenko}},
  \bibinfo{author}{\bibfnamefont{P.}~\bibnamefont{Wolynes}}, \bibnamefont{and}
  \bibinfo{author}{\bibfnamefont{H.}~\bibnamefont{Frauenfelder}},
  \bibinfo{journal}{J. Phys. Chem.} \textbf{\bibinfo{volume}{109}},
  \bibinfo{pages}{7488} (\bibinfo{year}{2005}).

\bibitem[{\citenamefont{Shenogina et~al.}(2008)\citenamefont{Shenogina,
  Keblinski, and Garde}}]{Shenogina2008}
\bibinfo{author}{\bibfnamefont{N.}~\bibnamefont{Shenogina}},
  \bibinfo{author}{\bibfnamefont{P.}~\bibnamefont{Keblinski}},
  \bibnamefont{and} \bibinfo{author}{\bibfnamefont{S.}~\bibnamefont{Garde}},
  \bibinfo{journal}{J. Chem. Phys.} \textbf{\bibinfo{volume}{129}},
  \bibinfo{pages}{155105} (\bibinfo{year}{2008}).

\bibitem[{\citenamefont{De~Los~Rios and Caldarelli}(2000)}]{DeLosRios2000}
\bibinfo{author}{\bibfnamefont{P.}~\bibnamefont{De~Los~Rios}} \bibnamefont{and}
  \bibinfo{author}{\bibfnamefont{G.}~\bibnamefont{Caldarelli}},
  \bibinfo{journal}{Phys. Rev. E} \textbf{\bibinfo{volume}{62}},
  \bibinfo{pages}{8449} (\bibinfo{year}{2000}).

\bibitem[{\citenamefont{Collet}(2001)}]{Collet2001}
\bibinfo{author}{\bibfnamefont{O.}~\bibnamefont{Collet}},
  \bibinfo{journal}{Europhys. Letters} \textbf{\bibinfo{volume}{53}},
  \bibinfo{pages}{93} (\bibinfo{year}{2001}).

\bibitem[{\citenamefont{Collet}(2011)}]{Collet2011}
\bibinfo{author}{\bibfnamefont{O.}~\bibnamefont{Collet}}, \bibinfo{journal}{J.
  Chem. Phys.} \textbf{\bibinfo{volume}{134}}, \bibinfo{pages}{085107}
  (\bibinfo{year}{2011}).

\end{thebibliography}

\end{document}